\documentclass[11pt,a4paper]{article}
\usepackage{amsfonts}
\usepackage{amssymb}
\usepackage{graphicx}
\usepackage{amsmath}
\usepackage[dvipdfm]{hyperref}
\usepackage{enumerate}
\RequirePackage{mathrsfs} \RequirePackage[sc]{mathpazo}
\RequirePackage{wasysym} \RequirePackage{setspace}

\makeatletter \@addtoreset{equation}{section}

\newcommand{\be}{\begin{equation}}
\newcommand{\ee}{\end{equation}}
\newcommand{\bea}{\begin{eqnarray}}
\newcommand{\eea}{\end{eqnarray}}
\begin{document}

\title{On  Hexagonal Structures in Higher Dimensional Theories }
\author{  Adil  Belhaj$^{1}$\thanks{belhaj@unizar.es},
Luis J. Boya$^{2}$\thanks{luis.boya@gmail.com}, Antonio  Segui$^{2}$\thanks{segui@unizar.es}\hspace*{-15pt} \\
\\
{\small $^{1}$ Centre of Physics and Mathematics, CPM-CNESTEN, Rabat, Morocco  }  \\
{\small $^{2}$ Departamento de  F\'isica Te\'orica, Universidad de Zaragoza, E-50009-Zaragoza, Spain}} \maketitle

\begin{abstract}
We analyze the geometrical background under which many Lie groups relevant to particle physics
are endowed with a (possibly multiple) hexagonal structure. There are several groups appearing, either as special holonomy groups on the compactification process from higher dimensions, or as dynamical string gauge groups; this includes groups like $\mbox{SU}(2)$, $\mbox{SU}(3)$, $G_2$,
$\mbox{Spin}(7)$, $\mbox{O}(8)$ as well as $E_8$ and $\mbox{O}(32)$. We emphasize also the
relation of these hexagonal structures with the octonion division algebra, as we expect as well eventually some role for octonions in the interpretation of symmetries in High Energy Physics.
\end{abstract}

\thispagestyle{empty} \newpage \setcounter{page}{1} \newpage

\section{Introduction}
Symmetry is an essential ingredient of almost all physical theories. In particular,
invariance under certain Lie groups characterizes e.g. Minkowski space and most of
interactions are regulated also by certain  gauge  groups, like $\mbox{SU}(3)$(color),
$\mbox{SU}(2)$ and $\mbox{U}(1)$ for the three forces of particle physics.

In this paper we remark the important r\^ole placed by hexagonal root structures in
higher dimensional theories, all related to superstrings. This comes about because several
compactification groups show in their root systems a hexagonal pattern. For such a group $G$,
we have a relation between the dimension and the rank which reads
\begin{equation}
\mbox{Dim}\;G = \mbox{rank} \;G + 6\times m,
\end{equation}
where $m$ is an integer specified later on.

The symmetry group is already the Lie group itself  $G$, but besides the Weyl discrete symmetry group of the Dynkin diagram stresses, in our case, the hexagonal symmetry. At the moment, we just want to state these peculiarities and symmetries, leaving for further work to grasp the true meaning of that coincidences in string theory and related models.

To motivate the study, we first consider the traditional superstring
theory, living in $10$ dimensions \cite{1}. If we impose ${\cal
N}=1$ supersymmetry in our realizable 4-dimensional world, we are
forced to compactify the remaining $10-4=6$  dimensions in a
Calabi-Yau threefold ($\mbox{CY}_3)$. The holonomy group \cite{2} of
a general $\mbox{CY}_n$  manifold is $\mbox{SU}(n)$, and in our
$n=3$ case it is reduced to  $\mbox{SU}(3)$. Its dimension  is
$\mbox{dim SU}(3)= 8 = 2+6= \mbox{ rank two} + \mbox{two simple
roots} + 4\;\mbox{reflections}$. In the corresponding Lie algebra
noted $A_2$, there are the two simple roots $\alpha_1$ and
$\alpha_2$ of equal length, and at $120^\circ$ angle. Its Dynkin
diagram is as follows
\begin{equation*}
\mbox{
         \begin{picture}(20,30)(70,0)
        \unitlength=2cm
        \thicklines
    \put(0,0.2){\circle{.2}}
     \put(.1,0.2){\line(1,0){.5}}
     \put(.7,0.2){\circle{.2}}
     \put(-1.2,.15){$A_{2}:$}
  \end{picture}
}  \label{ordAk}
\end{equation*}
The total six nonzero roots come from these simple ones, with the
sum and the opposed, to form the well known (single)  hexagon. See Fig.-1-
\begin{figure}[tbph]
\centering
\begin{center}
\includegraphics[width=8cm]{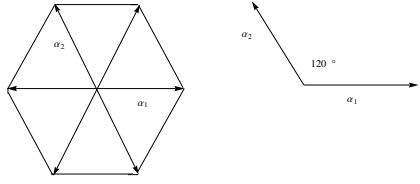} \caption{Root
system of $A_2$.
 }
\end{center}
\end{figure}

The Weyl group of the $A_2 (\equiv \mbox{su}(3))$  Lie algebra is
the symmetric group
    $S_3 = D_3 = Z_3 \rtimes Z_2$ of order $6$. It acts transitive and freely (= trivial
    stabilizer) on the hexagon.

 This appearance of  $\mbox{SU}(3)$ in string theory is the more remarkable, as it is the third
     time this group appears in fundamental physics. We remember the original group  $\mbox{SU}(3)$ flavour
      of Gell-Mann (1962) \cite{3}, mixing the first three quarks flavors (up $u$, down $d$ and strange $s$).
       Later (1972), Gell-Mann and Fritzsch introduced also $\mbox{SU}(3)$ as the gauge (color) group
        of the strong interactions\cite{4}, giving rise to QCD. At this point we only remark that the
         presence of the three $\mbox{SU}(3)$ groups in microphysics seem to be completely unrelated with each other.


\section{ Holonomy groups }
As we are to consider other holonomy groups, it  is good to recall here briefly Berger  classification
 of them, in this particular form, depending on the field of numbers and on some restrictions on
 the manifolds \cite{2}.
    We shall follow the approach in \cite{5}. A general (paracompact) $n$-dimensional
     manifold $\cal M$ admits always a Riemannian structure, whose generic holonomy group is
     the full orthogonal group $\mbox{O}(n)$. If $\cal M$  is just orientable, the group
     becomes the connected rotation (sub)group $\mbox{SO}(n)$; the condition for this is that
      the first Stiefel-Whitney class vanishes, $w_1(\mathcal{M}) = 0$. For complex  K\"{a}hler manifolds
       of complex dimension $n$, the game is played by the unitary $\mbox{U}(n) $ and the unitary
        unimodular (sub)group $\mbox{SU}(n)$, and again the obstruction is measured by the first
        Chern class $c_1(\cal M)$  of the (complex) tangent bundle. One completes the table with
         some quaternionic and octonionic groups. Here is the full classification of special holonomy
          groups \cite{2,5}, as related to the four division number algebras $\mathbb{R}, \mathbb{C}, \mathbb{H}$ and $\mathbb{O}$:

\begin{center}
\begin{tabular}{lll}
Number Field& General group & Unimodular Form \\
\hline
$\mathbb{R}$ & $\mbox{O}(n)$ & $\mbox{SO}(n)$ (orientable, $w_{1}=0$) \\
$\mathbb{C}$ & $\mbox{U}(n)$ (K\"{a}hler) & $\mbox{SU}(n)$ (Calabi-Yau, $c_{1}=0$) \\
$\mathbb{H}$ & $\mbox{q}(n)$ & $\mbox{Sq}(n)$ \\
$\mathbb{O}$ & $Spin(7)$ in $8d$ spaces & $G_{2}$ in $7d$ spaces
\end{tabular}
\end{center}

For $\mbox{q}(n)$ we mean the compact form of the non-simple group

\begin{equation}
\mbox{q}(n) = \frac{ \mbox{q}(1) \times \mbox{Sq}(n)}{Z_2}
\end{equation}
where $\mbox{q}(1) = \mbox{SU}(2)$ is the (multiplicative) group of
unit quaternions $\mathbb{H}$, which form the three-sphere $S^3$.
$\mbox{Sq}(n)$ is the intersection of the unitary group with the
real symplectic group, $\mbox{Sq}(n):= \mbox{U}(n) \cap
\mbox{Sp}(n)$. Today we have plenty of examples of manifolds with
any of these special holonomy groups \cite{6}. Notice all holonomy
 groups are compact, as subgroups of some real orthogonal group
 $\mbox{O}(n)$.\\
 As the octonion numbers $\mbox{O}$  are not associative  ($o_1(o_2o_3)\neq (o_1o_2)o_3$ in
    general)  only octonionic vector spaces could form up to three (octonionic) dimensions. In
    particular, the exceptional octonionic holonomy groups are only $\mbox{Spin}(7)$, with a
    natural irreducible (real) representation of dimension  $2^{(7-1)/2} = 8$, and the exceptional
     Lie group $G_2$, acting naturally in $\mathbb{R}^7$, i.e. in seven
     dimensions.  The corresponding  Dynkin diagrams are given by
\begin{figure}[tbph]
\centering
\begin{center}
\hspace{1.5cm} \includegraphics[width=8cm]{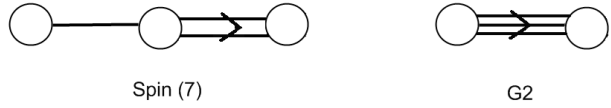}
\caption{Spin(7) and $G2$ Dynkin diagrams}
\end{center}
\end{figure}

We consider $G_2$ (with dimension $14$) as the automorphism group of
the octonions; if we write them as $o= v + \boldsymbol {\xi}$, with
$v$ the real part and $\boldsymbol {\xi}$ a vector in $R^7$; $v$ is
unchanged under any automorphism, as Aut(real numbers)=Id.
Therefore, as the norm $\mathcal{N}(o):=v^2+\boldsymbol{\xi} \cdot
\boldsymbol {\xi} $ is real, so is also invariant; so $G_2$ acts
naturally in the $6$-sphere of unit imaginary octonions; one shows
this action is transitive, so the stabilizer group (called \emph
{little group} in physics by Wigner) is of dimension $14-6=8$; it
turns out to be SU(3). Analogously, the $O(7)$ spin representation
$\Delta$ has dimension $\Delta= 2^{(7-1)/2} =8$ (and real type), and
the natural action of Spin(7) on the $7$-sphere is also transitive,
with stabilizer of dimension $7 \cdot 6/2 -7=14$; indeed it is the
group $G_2$ again (See \cite{17}). If one imagines this $7$-sphere
as the set of \underline {all} unit octonions, the imaginary ones
collect themselves in the equator $ \approx S^6$, and the mentioned
stabilizer SU(3) acts in the whole ambient space $R^6$ via the
$6=3+\bar{3}$ representations. This "octonionic" presentation of
SU(3) may very well be, on the long run, the very reason of the
presence so repeated of SU(3) in microphysics.

\section{ Hexagonal structures in M and F-theories}

In the "second" superstring revolution (1995), Witten \cite{7} showed how the five existing
superstring theories, living  in ten dimensions, were subsumed in a certain "M-Theory",
working in eleven dimensions, and encompassing maximal supergravity (32 supercharges).

As Townsend {\em et al.} stated in \cite{8} the corresponding seven-dimensional $(7= 11-4)$
compactifying manifold has most likely $ G_2$ holonomy and this group also exhibits a hexagonal root structure with rank $2$ and dimension $14$:
\begin{equation}
14 =2+ 6\times 2
\end{equation}
So we have two hexagons of unequal side length, with primary unequal
roots at angle $60^{\circ}+ 90^{\circ} = 150^{\circ}$. Two unequal roots each
generates an hexagon (we refrain to copy the star-shaped figure; see
\cite{9,10}). The Weyl group of the $G_2$ diagram is the dihedral
group
\begin{equation}
D_6 = Z_6\rtimes Z_2,
\end{equation}
of order 12 which acts transitively (stabilizer $Z_2$) on the two hexagons independently.

To see the nontriviality of our approach, notice e.g.
that for $B_2 = C_2$ does not work: $10 = 2 + 2\times 4$: no hexagons here! The figure is that of
 a square with diagonals. The two fundamental {\em{irreps}} of $G_2$, associated  with its  diagram
  are $7$ and $14$-dimensional (the first acts in unit imaginary octonions, as said, and the other is the adjoint).
   We wish to specify the action of these groups on spheres.

The group SU(3) has the complex representation $3$, and SU(3) being
unitary it leaves the $5$-sphere $S^5 \subset R^6 = C^3 $ invariant;
indeed, that action is again transitive, with little group SU(2):
$\mbox{SU(3)}/\mbox{SU(2)} \approx S^5$, or as an exact {\em suite}:
\begin{equation}
\mbox{SU(2)} \rightarrow \mbox{SU(3)} \rightarrow S^5
\end{equation}

On the other hand, as we explained above, $G_2 / \mbox{SU(3)}
\approx S^6$ also, so we have another exact sequence
\begin{equation}
\mbox{SU(3)} \rightarrow G_2 \rightarrow S^6
\end{equation}

The IIB superstring theory did not fit very well in the original
M-Theory, so C. Vafa postulated \cite{11}, (1996) a further
extension to 12 dimensions. Today this is an ample field of
research, still growing, \cite{12}; for more work see e.g. the
recent F-theory paper in relation to GUT \cite{13}. Working with
F-theory, we need  a $ 12 - 4 = 8$ dimensional compactifying
manifold. Candidate groups are $\mbox{ O}(8)$,  $\mbox{SO}(8)$,
$\mbox{Spin}(7)$ and $\mbox{SU}(4)$, all with a natural real 8-dim
representation. $\mbox{Spin}(7)$ is the natural generalization of
$G_2$, as it can be understood also as an octonionic group. It has a
natural spinor $ 8 = 2^{(7-1)/2}$-dimensional real representation,
therefore it acts naturally in the seven sphere of all unit
octonions. It is also one of the holonomy groups in Berger's list,
see above. The O(7) group has dimension $ 7 \cdot 6 /2 = 21$, and
rank three: it is $B_3 \neq C_3$ in Cartan's list; hence, the number
of nonzero roots is
\begin{equation}
(Dim G - rank G) = (21-3) = 18 = 6 \cdot 3,
\end{equation}
so it does have an hexagonal structure.

This is seen also from the Dynkin diagram $B_3$: {\setlength{\unitlength}{1mm}
\begin{picture}(23,15)
    \put(0,0){\circle{2}}
    \put(10,0){\circle{2}}
    \put(20,0){\circle*{2}}


    \put(1,0){\line(1,0){8}}

    \multiput(10,-1)(0,2){2}{\line(1,0){10}}



\end{picture}}, interpreted as the Weyl symmetry group. This Weyl group $ W(B_3) = Z_2^3 \rtimes S_3$, of order $48$, has three involutive generators, say $a$, $b$, $c$ with relations
\begin{equation}
a^2 = b^2 = c^2 = (ab)^3 = (bc)^4 = (ac)^2 = e
\end{equation}
The biplane $<a,b>$ generates a hexagon as before (group $\mbox{SU}(3))$. The biplane $<b,c>$
generates a square (with diagonal and sides), with symmetry group $D_4 = Z_4 \rtimes Z_2$, and
the biplane $<a,c>$ with orthogonal fundamental roots generates an unequal cross, with group
$V = Z_2\times Z_2$. The counting of nonzero roots is
\begin{equation}
18 =21-3  = 6 (\mbox{hexagon}) + 8 (\mbox{diagonalized square}) + 4 (\mbox{unequal cross})
\end{equation}
Thus we see here again a hexagon emerging.

So the three standard compactifying groups show hexagons: $\mbox{Spin}(7)$, $ G_2$ and $\mbox{SU}(3)$, and the three are also related to the octonions. To call attention to this fact is the main purpose of this paper.

As $\mbox{Spin}(7)$ lies naturally in $\mbox{SO}(8)$, which acts
also naturally in $S^7$, it is tempting to amplify the scheme
including $\mbox{SO}(8)$. The Dynkin diagram for $D_4$ is very
special (Figure 3). \vspace{1cm}
\begin{figure}[tbph]
\centering
\begin{center}
\hspace{1.5cm} \includegraphics[width=12cm]{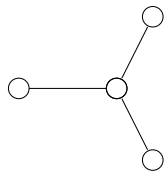} \caption{$D_4$
\, Dynkin diagram}
\end{center}
\end{figure}

Notice that the $D_4$ Dynkin Diagram  is the only diagram with outer
symmetry beyond $Z_2$, as in the $A$  and  $D$ series and $E_6$. The
three external roots correspond to the three dim 8 {\em{irreps}}:
vector and the two spinor $\Delta_{L, R}$, permuted by triality, as
there is a large group of outer automorphisms:
\begin{equation}
\mbox{Out(Spin}(8)) = S_3
\end{equation}
 We believe this triality would play a r\^ole in understanding the hexagon replicas (and hopefully
  other features of the physics), but at the moment we do not elaborate.

We just comment an aspect of this diagram.  The central root (which corresponds to the adjoint
{\it{irrep}}, of dimension  28) with any of the 3 outer ones generates {\it three} hexagons, as each has the structure of $A_2$ .  The three outside roots by themselves generate (as involutions)
$(Z_2)^3$, as they are disconnected and hence orthogonals. In general, the hexagon structure arises
because the numerical coincidence  $ \mbox{Dim}\; G = \mbox{rank} + 6\times m$, where $m$ is now 4:
\begin{equation}
\mbox{Dim O}(8) = 8\times 7/2 = 28 = 4 \mbox{(rank)} + 6\times 4.
\end{equation}

Now the nine mentioned groups can be assembled together after the inclusions

\begin{equation}
\begin{tabular}{cccccccccc}
Group:& SU(2)&$\subset$& SU(3) &$\subset$&  G$_2$ &$\subset$ & Spin(7)& $\subset$ & SO(8) \\
Dim   & 3    &         &   8   &         &   14   &          &   21   &           &  28   \\
\end{tabular}
\end{equation}
Together with the natural ones
\begin{equation}
\begin{aligned}
&\text{SU(2)}\subset \text{SU(2)}^2=\text{Spin(4)}\rightarrow \text{SO(4)} \\
&\text{Spin(7)}\supset \text{Spin(5)} \\
&\text{SU(4)} \subset \text{SO(8)} \supset \text{SU(7)}
\end{aligned}
\end{equation}

>From the Dynkin diagram for $D_4$, one gets by truncation
(removing a node) and folding (collapsing under external automorphisms) all the groups above;
 it is a standard exercise in Lie groups.
Notice also the equivalences (or Cartan identities)
\begin{eqnarray}
 A_1 + A_1 &=& D_2  (\mbox{group O}(4)),\;\; \mbox{or\; Spin}(4) = \mbox{(Spin}(3))^2\\
A_3 &=& D_3, \;\;\mbox{or \; SU}(4) = \mbox{Spin}(6)\nonumber.
\end{eqnarray}
By folding we get  $B_2 =\mbox{Spin}(5) = \mbox{Sp}(2) = C_2$.\\
Now we see  the "6'' number for all holonomy groups. The following
Table sums up the situation:
 \begin{center}
\begin{tabular}{lll}
Holonomy  group &  Dimension  &   Decomposition\\
\hline
O(8) & 28& 4+6$\times$4 \\
Spin(7) & 21 & 3+6$\times$3 \\
 $G_2$ & 14 & 2+6$\times$2\\
SU(4) & 15 & 3+6$\times$2\\
SU(3) & 8 & 2+6$\times$1,
\end{tabular}
\end{center}
\bigskip
We see that all include hexagons.

\section{Discussion}
It seems remarkable to us that even the large groups used as gauge
groups in string theory,  and related  models,  can group the
nonzero roots in some hexagons.
$E_8$  appears as itself in M-theory \cite{15}, and as squared $E_8\times E_8$ in one of the heterotic strings.
But, in dimensions $248$ and $\mbox{rank}(8)$  we have again nonzero roots grouped in hexagons: in dealing with
simply laced groups, there is an hexagon for each $A_2$  bond  for example, in $E_8$
\begin{equation}
\mbox{Dim}\; E_8 =248= 8 (\mbox{rank}) + 6\times 40.
\end{equation}
As for the other heterotic string group, $\mbox{SO}(32)$, it is also a $D$ series groups, which is
simply laced one, and it  should include hexagons. Indeed, we have
\begin{equation}
\mbox{ Dim SO(32)}=32\times 31/2 = 496 = \mbox{third perfect number}
= 16(\mbox{rank}) + 6\times 80.
\end{equation}
We realize it is hard to imagine any role for these numerous
hexagons, but the numbers are there.  Also, at face value we do not
see any r\^ole for the octonions.

We cannot skip the appearance of hexagonal structures in other domains of physics. We add this just to reinforce our point of view; in particular hexagons, dual triangles and squares tessellate the plane $R^2$. For another example, in solid state physics, the structure of graphene \cite{16} (or laminar graphite) has been shown a hexagonal structure, linked to the benzene hexagon.  One also expects some hexagonal structures to arise also in valence-four elements, as carbon; in particular, silicium(Si) and germanium(Ge) are expected to have hexagonal structures also.

\emph{Acknowledgments:} AB would like to thank the  grant
A/031268/10  and the Department of Theoretical Physics at Zaragoza
University for very  kind hospitality. He is  also very grateful to
M. Asorey for  scientific help. This work has been also supported by
the grants FPA2009-09638 (CICYT), 2011-E24/2 (DGHD-DGA) and A/031268/10.

\end{document}